\documentstyle[psfig]{mn}
\begin{document}
\textheight 8.5 in
\date{}
\title[Non-linear density evolution in spherical collapse model]{Non-linear density evolution from an  improved spherical collapse
model}
\author[Sunu Engineer, Nissim Kanekar and T. Padmanabhan]
{Sunu Engineer $^{1}$ \thanks{sunu@iucaa.ernet.in}, Nissim Kanekar$^{2}$ \thanks{nissim@ncra.tifr.res.in} and T. Padmanabhan$^{1}$\thanks{paddy@iucaa.ernet.in}\\
$^{1}$Inter-University Centre for Astronomy and Astrophysics,\\
 Post Bag 4, Ganeshkhind, Pune 411 007, India \\
$^{2}$National Centre for Radio Astrophysics,\\
 Post Bag 3, Ganeshkhind, Pune 411 007, India }

\maketitle

\begin{abstract}
\noindent We investigate the evolution of non--linear density perturbations
by taking into account the effects of deviations from spherical symmetry 
of a system.  Starting from the standard spherical top hat model in which 
these effects are ignored, we introduce a physically motivated closure condition 
which specifies the dependence of the additional terms on the density contrast, 
$\delta$. The modified equation can be used to model the behaviour of an overdense 
region over a sufficiently large range of $\delta$. The key new idea is 
a Taylor series expansion in ($1/\delta$) to model the non--linear epoch. 
We show that the modified equations quite generically lead to the formation 
of stable structures in which the gravitational collapse is halted at around 
the virial radius. The analysis also allows us to connect up the behaviour of 
individual overdense regions with the non--linear scaling relations satisfied 
by the two point correlation function.

\end{abstract}

\begin{keywords}
Cosmology : theory -- dark matter, large scale structure of 
the Universe
\end{keywords}

\section{Introduction}
\label{sec:Intro}
\noindent  Analytic modelling of the non--linear phase of gravitational clustering
has been a challenging but interesting problem upon which a considerable
amount of attention has been bestowed in recent years. The simplest, yet
remarkably successful, model for non-linear evolution is the 
 Spherical Collapse Model (SCM, hereafter), which has been  applied in
the study  of  various empirical results in the gravitational instability 
paradigm.  Unfortunately, this approach has  serious flaws --- both 
mathematically and conceptually. Mathematically, the SCM has a singular 
behaviour at finite time and predicts infinite density contrasts for all 
collapsed objects. Conceptually, it is not advisable to model the real 
universe as a sphere, in spite of the standard temptations to which 
theoreticians often succumb. The two issues are, of course, quite 
related, since, in any realistic situation, it is the deviations from spherical 
symmetry which lead to virialised stable structures getting formed. 
In conventional approaches, this is achieved by an {\it ad hoc\/} method
which involves halting the collapse at the virial radius by hand and mapping the
resulting non-linear and linear overdensities to each other. This leads to
the well known rule-of-thumb  that, when the linear overdensity is about 
$1.68$, bound structures with non-linear overdensities of about $178$ would 
have formed. The singular behaviour, however, makes the actual trajectory
of a spherical system quite useless after the turnaround phase ---
a price we pay for the arbitrary procedure used in stabilizing the system. 
But the truly surprising feature is that, despite its inherent arbitrariness, 
the SCM, when properly interpreted, seems to give useful insights into the behaviour 
of real systems. The Press--Schechter formalism (Press \& Schechter 1974), for 
the abundance of bound structures, uses SCM implicitly; more recently, it was
shown that the basic physics behind the non-linear scaling relations (NSR)
obeyed by the two point correlation function can be obtained from a judicious
application of SCM (Padmanabhan 1996a). These successes, as well as the 
inherent simplicity of the underlying concepts, make the SCM an attractive 
paradigm for studying non-linear evolution in gravitational clustering and 
motivate one to ask : Can we improve the basic model in some manner so that 
the behaviour of the system after turnaround is `more reasonable' ?
\par
\noindent It is clear from very general considerations that such an approach has
to address  fairly non-trivial technical issues. To begin with, 
{\it exact } modelling of deviations from spherical symmetry is quite
impossible since it essentially requires solving the full BBGKY hierarchy.
Secondly, the concept of a radius $R(t)$ for a shell, evolving
only due to the gravitational force of the matter inside, becomes ill-defined
when deviations from spherical symmetry are introduced. Finally, our real 
interest is in modelling the statistical features of the density growth; 
whatever modifications we make to SCM should eventually tie up with known 
results for the evolution of, for instance, the two point correlation function. 
That is, we have to face the question of how best to obtain the {\it statistical\/} 
properties of the density field from the behaviour of a {\it single\/} system.
\par
\noindent In this paper, we try to address these problems in a limited but focussed manner. 
We tackle the deviations from spherical symmetry by  the retention of a 
term (which is usually neglected) in the equation describing the growth of the 
density contrast. Working in the fluid limit, we show that 
this term is physically motivated and  present some arguments  to derive 
an acceptable form for the same. The key new idea is to introduce a Taylor 
series expansion in ($1/\delta$) (where $\delta$ is the density contrast) to 
model the non-linear evolution. We circumvent the question of defining the
`radius' of the non-spherical regions by working directly with density
contrasts. Finally, we attempt to make the connection with statistical descriptors
of non-linear growth, by using the non-linear scaling relations known from 
previous work. More precisely, we show that the modified equations predict a 
behaviour for the relative pair velocity (when interpreted statistically) which 
agrees with the results of N-body simulations.
\par
\noindent The paper is divided into the following sections. The relevant equations
 describing the SCM are set out in Section 2; we also summarise the physical 
and {\it ad hoc\/} aspects of the SCM here. Next, we recast the equations in a
different form and introduce two functions (i) a ``virialization term''  and (ii) 
a function $h_{\rm SC}(\delta)$, whose asymptotic forms are easy to determine. 
The behaviour of $h_{\rm SC}(\delta)$ 
in the presence and absence of the ``virialization term'' is also detailed here. 
In Section 4,  we present the  arguments that give the functional  
forms for the above term over a large range of $\delta$; we then go on to present 
the results in terms of a single collapsing body and show how 
this term stabilizes a collapse which would have otherwise 
ended up in a singularity in terms of the growth of the density contrast 
with time. When this term is carried through into the equation for $R(t)$ for 
a single system, it can be seen the radius reaches a maximum and gracefully 
decreases to a constant, remaining so thereafter. In the standard SCM, the radius 
decreases from the maximum all the way down to zero, thereby causing the 
density to diverge. Section 5 summarises the results and discusses their 
implications. 

\section{The Spherical Collapse Model.}
\noindent The scales of interest in the current work are much smaller than 
the Hubble length and the velocities in question are non-relativistic; Newtonian 
gravity can hence be used for the following analysis. We will consider the case 
of a dust-dominated, $\Omega = 1$ Universe and treat the system in the fluid limit as 
being made up of pressureless dust of dark matter, with a smoothed density, 
$\rho_m(t,{\bf x})$, and a mean velocity, ${\bf v}(t,{\bf x})$. (This approach, 
of course, ignores effects arising from shell crossing and multi-streaming; these 
will be commented on later.)\\
\noindent The density contrast, $\delta({\bf x},t)$ is defined by \\
\begin{equation}
\rho_m(t,{\bf x}) = \rho_b(t) [1 + \delta({\bf x},t)]
\end{equation}
\noindent where $\rho_b$ denotes the smooth background density of matter. We 
define a velocity field $u^i = v^i/(a \dot{a})$, where $v^i$ is the peculiar 
velocity (obtained after subtracting out the Hubble expansion) and $a(t)$ 
denotes the scale factor. Taking the divergence of the field $u^i$ and
writing it as \\
\begin{equation}
\partial_{i} u_{j}=\sigma_{ij}+\epsilon_{ijk} \Omega^{k}+\frac{1}{3}
\delta_{ij} \theta
\end{equation}
\noindent where $\sigma_{ij}$ is the shear tensor, $\Omega^k$ is the rotation 
vector and $\theta$ is the expansion, we can manipulate the fluid equations (see
e.g. Padmanabhan 1996b) to obtain the following equation for $\delta$ \\
\begin{eqnarray}
\label{deltaequation}
\frac{d^2\delta}{d a^2}+\frac{3}{2 a}\frac{d \delta}{d a}-\frac{3}{2 a^2} 
\delta (1+\delta)\quad = \quad\qquad\qquad\qquad\qquad\nonumber \\
\qquad\qquad\qquad\frac{4}{3} \frac{1}{(1+\delta)} 
\left( \frac{d \delta}{d a}\right)^2+(1+\delta) (\sigma^2-2\Omega^2)
\end{eqnarray}
\noindent The same equation can be written in terms of time $t$ as \\ 
\begin{eqnarray}
\label{deltatime}
\ddot{\delta}-\frac{4}{3} \frac{\dot{\delta}^2}{(1+\delta)}+\frac{2
\dot{a}}{a} \dot{\delta} \quad=\quad\qquad\qquad\qquad\qquad\qquad\qquad\nonumber \\
\qquad\qquad\qquad\qquad 4 \pi G \rho_{b} \delta (1+\delta)+\dot{a}^2
(1+\delta) (\sigma^2-2 \Omega^2)
\end{eqnarray}
\noindent This equation turns out to be the same as the one for density contrast
in the SCM, except for the additional term \hskip 1 in $(1+\delta)(\sigma^2-2\Omega^2)$, 
arising from the angular momentum  and shear of the system. To see this explicitly,
 we introduce a function $R(t)$ by the definition
\begin{equation}
\label{deltadefn}
1+\delta= {{9GM{t^2}}\over {2R^3}} \equiv \lambda \frac{a^3}{R^3}
\end{equation}
\noindent where $M$ and $\lambda$ are constants. Using this relation 
between $\delta$ and $R(t)$, equation (\ref{deltatime}) can be converted 
into the following equation for $R(t)$ 
\begin{equation}
\label{reqn}
\ddot{R}=-\frac{GM}{R^2}-\frac{1}{3} \dot{a}^2 \left(
\sigma^2-2\Omega^2\right) R 
\end{equation}
\noindent where the first term represents the gravitational attraction
due to the mass inside a sphere of radius $R$  
and the second gives the effect of the shear and angular momentum. \\
\noindent In the case of spherically symmetric evolution, the shear and 
angular momentum terms can be set to zero; this gives \\
\begin{equation}
\label{standardSCM}
\frac{d^2 R}{d t^2}=-\frac{GM}{R^2}
\end{equation}
\noindent which governs the evolution of a spherical shell 
of radius $R$,  collapsing under its own gravity; $M$ can now be identified 
with the mass contained in the shell; this is standard SCM.\\
\par
\noindent At this point, it is important to note a somewhat subtle aspect of these
equations. The original fluid equations are clearly Eulerian in nature: 
{\it i.e.} the time derivatives give the temporal variation of the quantities  
at a fixed point in space. However, the time derivatives in equation 
(\ref{deltatime}), for the density contrast $\delta$, are of a different kind. 
Here, the observer is moving with the fluid element and 
hence, in this, Lagrangian case, the variation in density contrast seen 
by the observer has, along with the intrinsic time variation, a component 
which arises as a consequence of his being 
at different locations in space at different instants of time. When the 
$\delta$ equation is converted into an equation for the function $R(t)$, 
the Lagrangian picture is retained; in SCM, we can  interpret $R(t)$ as 
the radius of a spherical shell, co--moving with the observer. The mass 
$M$ within each shell remains constant in the absence of shell crossing 
(which does not occur in the standard SCM for reasonable initial
conditions) and the entire formalism is well defined. The physical 
identification of $R$ is, however, not so clear in the case where the 
shear and rotation terms are retained, as these terms break the spherical 
symmetry of the system. We will nevertheless continue to think of $R$ 
as the ``effective shell radius`` in this situation, {\it defined  by\/} 
equation (\ref{deltadefn}) governing its evolution. Of course, there is 
no such ambiguity in the {\it mathematical} definition of $R$ in  this formalism.\\
\par 
\noindent Before proceeding further, let us briefly summarize the results of 
standard SCM. Equation (\ref{standardSCM}) can be integrated to obtain $R(t)$ in the 
parametric form 
\begin{eqnarray}
\label{stheqn1}
R&=&\frac{R_i}{2 \delta_i} (1- \mbox{cos}\;\theta)\\
\label{stheqn2}
t&=& \frac{3 t_i}{4 \delta_i^{3/2}} (\theta-\mbox{sin}\;\theta)
\end{eqnarray}
\noindent where $R_i$, $\delta_i$ and $t_i$ are the initial radius, initial 
density contrast and initial time, respectively, with 
$R_i^3=(9GMt_i^2/2)(1+\delta_i)^{-1} \simeq (9GMt_i^2/2) $ for $\delta_i \ll 1$. 
Given $M$, there are only two independent constants, {\it viz\/} $t_i$ and $\delta_i$.
All the physical features of the SCM can be easily derived from the above
 solution. Each spherical shell expands at a progressively slower rate against the 
self-gravity of the system, reaches a maximum radius and then collapses under its 
own gravity, with a steadily increasing density contrast. The maximum radius, 
$R_{max}=R_i/\delta_i$, achieved by the shell,  occurs at a density 
contrast $\delta =(9\pi^2/16)-1 \approx 4.6$, which is in the ``quasi-linear'' 
regime. In the case of a perfectly spherical system, there exists no 
mechanism to halt the infall, which proceeds inexorably towards a 
singularity, with all the mass of the system collapsing to a single point. 
Thus, the fate of the shell (as described by equations (\ref{stheqn1}) and 
(\ref{stheqn2})) is to collapse to zero radius at $\theta = 2\pi$ with an infinite 
density contrast; this is, of course, physically unacceptable.
\par
 \noindent In real systems, however, the implicit assumptions 
that  (i) matter is distributed in spherical shells and (ii) the non-radial 
components of the  velocities of the particles are small, will 
break down  long before infinite densities are reached.
Instead, we expect the collisionless dark matter to reach virial equilibrium. 
After virialization, $|U|=2 K$, where $U$ and $K$ are, respectively, the potential 
and kinetic energies; the virial 
radius can be easily computed to be half the maximum radius reached by the system. 
\par
\noindent The virialization argument is clearly physically well-motivated for real systems. 
However, as mentioned earlier, there exists no mechanism in the standard SCM 
to bring about this virialization; hence, one has to
introduce  by hand the assumption  that, as the 
shell collapses and  reaches a particular radius, 
say $R_{max}/2$, the collapse 
is halted and the shell remains at this radius thereafter. This arbitrary 
introduction of virialization is clearly one of the major drawbacks of the standard
SCM and takes away its predictive power in the later stages of evolution.  We 
shall now see how the retention of the angular momentum  
term in equation (\ref{reqn}) can serve to stabilize the collapse of the system, 
thereby allowing us to model the evolution towards $r_{vir}=R_{max}/2$ smoothly.
\section{The $h_{\rm SC}(\delta)$ function.}
\label{sec:hfunction}
\noindent As detailed in the previous section, the primary defect of  
the standard SCM is the {\it ad hoc\/} nature of the stabilization of the shell 
against its collapse under gravity, which arises on account of 
the assumption of perfect spherical symmetry, implicit in the neglect of the shear 
and angular momentum terms. We hence return to equation (\ref{deltaequation}), 
retain the above terms, and recast the equation into a form more 
suitable for analysis. Using logarithmic variables, $D_{\rm SC} \equiv {\rm ln}
\hskip 0.03 in (1 + \delta)$ and $\alpha \equiv {\rm ln}\hskip 0.03 in a$, equation 
(\ref{deltaequation}) can be written in the form (the subscript `SC'
stands for `Spherical Collapse')
\begin{eqnarray}
\label{deltalog}
\frac{d^2 D_{\rm SC}}{d \alpha^2}-\frac{1}{3} \left(\frac{d D_{\rm SC}}{d
\alpha }\right) ^2 + \frac{1}{2} \frac{d D_{\rm SC}}{d \alpha} \quad = 
\qquad\qquad\qquad \quad\nonumber \\
\qquad\qquad\qquad\qquad \frac{3}{2} \left[\exp (D_{\rm SC})-1 \right] + a^2 (\sigma^2-2 \Omega^2)
\end{eqnarray} 
\noindent It is convenient to  introduce the quantity, $S$, defined by \\
\begin{equation}
S \equiv a^2 (\sigma^2-2 \Omega^2)
\end{equation}
\noindent  which we shall hereafter call the ``virialization term''.  The
 consequences of the retention of the virialization term are easy to
describe qualitatively. We expect  the 
evolution of an initially spherical shell to proceed along the lines of the standard SCM 
in the initial stages, when any deviations from spherical symmetry, present in the 
initial conditions, are small. However, once the maximum radius is reached and the 
shell recollapses, these small deviations are amplified by a positive feedback 
mechanism. To understand this, we note that all particles in a given spherical 
shell are equivalent due to the spherical symmetry of the system. This implies 
that the motion of any particle, in a specific shell, can be considered 
representative of the motion of the shell as a whole. Hence, the behaviour of the 
shell radius can be understood by an analysis of the motion of a single particle. 
The equation of motion of a particle in an expanding universe can be written as 
\begin{equation}
\ddot{{\bf X}_i}+2\frac{\dot{a}}{a} \dot{{\bf X}_i}=-\frac{\nabla \phi}{a^2}
\end{equation}
\par
\noindent where $a(t)$ is the expansion factor of the locally overdense ``universe".
The $\dot{{\bf X}_i}$ term  acts as a damping force when it is positive; 
{\it i.e.} while the background is expanding. However, when the
overdense region reaches the point of maximum expansion and turns around, this 
term becomes negative, acting like a {\it negative\/} damping
term, thereby amplifying any deviations from spherical symmetry 
which might have been initially present. Non-radial components of velocities 
build up, leading to a randomization of velocities which finally results 
in a virialised structure, with the mean relative velocity between any 
two particles balanced by the Hubble flow. It must be kept in mind, 
however, that the introduction of the virialization term  changes the 
behaviour of the solution in a global sense and it is  not strictly 
correct to say that this term starts to play a role {\it only after}
  recollapse, with the evolution proceeding along the lines of the 
standard SCM until then. It is  nevertheless reasonable to expect that, 
at early times when the term is small, the system will evolve as standard SCM  
 to reach a maximum radius, but will fall back smoothly to a constant size  later on. 
\par
\noindent The virialization term, $S$, is, in general, a function of $a$ and ${\bf x}$, especially 
since the derivatives in equation (\ref{deltatime}) are total time derivatives, 
which, for an expanding Universe, contain partial derivatives with respect 
to both ${\bf x}$ and $t$ separately.  
Handling  this equation exactly will take us back to the full non-linear equations for the
fluid and, of course, no progress can be made. Instead, we will make the
 {\it ansatz\/}   that the virialization term depends on $t$ and ${\bf x}$
only through $\delta(t,{\bf x})$.
\begin{equation}
S(a,{\bf x}) \equiv S(\delta(a,{\bf x})) \equiv S(D_{\rm SC})
\end{equation}
\noindent In other words, $S$ is a function of the density contrast alone. 
This {\it ansatz\/}  seems well  motivated because  the density contrast, $\delta$,
 can be used to characterize the SCM at any point in its evolution and one might 
 expect the virialization  term to be a function only of the system's state, at
least to the lowest order. Further, the results obtained with this assumption 
appear to be sensible and may be treated as a test of the {\it ansatz\/} in its 
own framework.\\ 
\noindent To proceed further systematically, we {\it define} a function $h_{\rm SC}$
by the relation \\
\begin{equation}
\label{defh}
{{dD_{\rm SC}}\over {d\alpha}} = 3h_{\rm SC}
\end{equation}

\noindent For consistency, we  shall assume the {\it ansatz\/}  $h_{\rm SC}(a,{\bf x}) \equiv
 h_{\rm SC}\left[\delta(a,{\bf x})\right]$.
The definition of $h_{\rm SC}$ allows us to write equation (\ref{deltalog}) as \\
\begin{equation}
\label{hequation}
\frac{d h_{\rm SC}}{d \alpha}=h_{\rm SC}^2-\frac{h_{\rm SC}}{2}+\frac{1}{2} 
\left[\exp (D_{\rm SC}) -1\right] + \frac{S(D_{\rm SC})}{3}
\end{equation}
\noindent Dividing (\ref{hequation}) by (\ref{defh}), we obtain the following 
equation for the function $h_{\rm SC}(D_{\rm SC})$\\
\begin{eqnarray}
\label{dhdDeqn}
\frac{dh_{\rm SC}}{dD_{\rm SC}} = \frac{h_{\rm SC}}{3}-\frac{1}{6}+ \frac{1}{6 h_{\rm SC}}
\left[\exp(D_{\rm SC})-1\right]+\frac{S(D_{\rm SC})}{9 h_{\rm SC}}
\end{eqnarray}
\par
\noindent If we know the form  of either $h_{\rm SC}(D_{\rm SC})$ or $S(D_{\rm SC})$,
this equation allows us to determine the other. Then, using equation (\ref{defh}),
one can determine $D_{\rm SC}$. Thus, our modification of the standard SCM 
essentially involves providing the form of $S_{\rm SC}(D_{\rm SC})$ or 
$h_{\rm SC}(D_{\rm SC})$. 
We shall now discuss several features of such a modelling in order to arrive 
at a suitable form.\\
\par 
\noindent The behaviour of $h_{\rm SC}(D_{\rm SC})$ can be qualitatively understood from 
our knowledge of the behaviour of $\delta$ with time. In the linear regime 
($\delta \ll 1$), we know that $\delta$ grows linearly with $a$; hence 
$h_{\rm SC}$ increases with $D_{\rm SC}$. At the extreme non-linear end ($\delta \gg 1$), 
 the system ``virializes'', {\it i.e.\/}  the proper radius and the density of the system become
constant. On the other hand, the density $\rho_b$, of the background, falls like $t^{-2}$ 
(or $a^{-3}$) in a flat, dust-dominated Universe. The density contrast  
is defined by $\delta = (\rho/\rho_b - 1) \sim \rho/\rho_b$ (for $\delta \gg 1$) 
and hence \\
\begin{equation}
\delta \propto t^2 \propto a^3
\end{equation}
\noindent in the non-linear limit. Equation (\ref{defh}) then implies that 
$h_{\rm SC}(\delta)$ tends to unity for $\delta \gg 1$. Thus, we expect that 
$h_{\rm SC}(D_{\rm SC})$ will start with a value far less than unity, grow, reach a 
maximum a little greater than one and then smoothly fall back to unity. 
[A more general situation  discussed in the literature corresponds to $h 
\rightarrow {\rm constant}$ as $\delta \rightarrow \infty$, though the 
asymptotic value of $h$ is not necessarily unity. Our discussion can be 
generalised to this case and we plan to explore this in a future work.]
\par
\noindent This behaviour of the $h_{\rm SC}$ function can be given another useful 
interpretation whenever the density contrast  has a monotonically 
decreasing relationship with the scale, $x$, with small $x$ implying large 
$\delta$ and vice-versa. Then, if we use a local power law approximation 
$\delta \propto x^{-n}$ for $\delta \gg 1$ with some $n >0$, $D_{\rm SC} 
\propto \ln (x^{-1})$ and 
\begin{equation}
h_{\rm SC} \propto {{dD_{\rm SC}} \over 
{d\alpha}} \propto - {{{d \ln} ({1\over x})}\over {d \ln a}} \propto 
\frac{\dot{x} a}{\dot{a} x} \propto - {v \over {{\dot a}x}}
\end{equation}
\par
\noindent where $v \equiv a{\dot x}$ denotes the mean relative velocity.
Thus, $h_{\rm SC}$ is proportional  to the ratio of the peculiar velocity 
to the Hubble velocity. We know that this ratio is small 
in the linear regime (where the Hubble flow is dominant) and later 
increases, reaches a maximum and finally falls back to unity with the 
formation of a stable structure; this is another argument leading to
the same  qualitative behaviour 
of the $h_{\rm SC}$ function.
\par
\noindent Note that, in standard SCM (for which $S = 0$), equation 
(\ref{dhdDeqn}) reduces to \\
\begin{equation}
\label{dhdDscm}
3h_{\rm SC}\frac{dh_{\rm SC}}{dD_{\rm SC}}=h_{\rm SC}^2-{h_{\rm SC}\over 2}+{\delta \over 2}
\end{equation}
\noindent The presence of the linear term in $\delta$ on the RHS of the 
above equation causes $h_{\rm SC}$ to increase with $\delta$, with $h_{\rm SC} \propto 
\delta^{1/2}$ for $\delta \gg 1$. If virialization is imposed as  
an {\it ad hoc\/}  condition, 
then  $h_{\rm SC}$ should fall back to unity discontinuously --- which is 
clearly unphysical; the form of $S(\delta)$ must hence be chosen so as to ensure 
a smooth transition in $h_{\rm SC}(\delta)$ from one regime to another.
\par
\noindent As an aside, we would like to make some remarks on the nature of the
``virialization term'', $S(\delta)$, in a somewhat wider context. As is well-known, 
gravitational clustering can be described at three different levels of 
approximation, by different mathematical techniques. The first
approach tracks the clustering by following the true particle trajectories;
this is what is done, for example, in N-body simultations. This method does
not involve any approximation (other than the validity of the 
Newtonian description at the scales of interest); it is, however, clearly analytically 
intractable.
At the next level, one may describe the system by an one-particle distribution funtion
and attempt to solve the collisionless Boltzmann equation for
the distribution function $f(t,{\bf x},{\bf v})$; the approximation here lies in the
neglect of gravitational collisions, which seems quite reasonable as the time scale
for such collisions is very large for standard dark matter particles. Finally, 
one can treat the system in the fluid limit described by five functions:
the density $\rho(t,{\bf x})$, mean velocity ${\bf v}(t,{\bf x})$,
and gravitational potential $\phi(t,{\bf x})$, thus neglecting multi-streaming
effects. Our analysis was based on this level of approximation. The key difference 
between the last two levels of description lies in the fact that the distribution function
allows for the possibility of different particle velocities at any point in space
({\it i.e.} the existence of {\it velocity dispersions}), while the fluid
picture assumes a mean velocity at each point. It is also known
that the gradients in velocity dispersion can provide a kinetic pressure
which will also provide support against gravitational collapse.
While a detailed analysis of these terms is again exceedingly difficult, one can
incorporate the lowest order effects of the gradient in the velocity dispersion by
modifying equation (\ref{deltalog}) to the form \\
\begin{eqnarray}
\label{new_deltalog}
\frac{d^2 D_{\rm SC}}{d \alpha^2}-\frac{1}{3} \left(\frac{d D_{\rm SC}}{d
\alpha }\right) ^2+\frac{1}{2} \frac{d D_{\rm SC}}{d \alpha} \quad = \qquad
\qquad\qquad\nonumber \\
\qquad\qquad\frac{3}{2} \left[\exp (D_{\rm SC})-1 \right] + 
a^2 (\sigma^2-2 \Omega^2) + f(a,x)
\end{eqnarray}
\noindent where $f(a,x)$ contains the lowest order contributions from the dispersion terms. We
can then define \\
\begin{equation}
S(a,x) = a^2 (\sigma^2-2 \Omega^2) + f(a,x)
\end{equation}
and again invoke the {\it ansatz} $S(a,x) \equiv S(\delta) $. Note that
$S(\delta)$ now contains the lowest order contributions arising from shell crossing,
multi-streaming, etc., besides the shear and angular momentum terms, {\it
i.e.} it contains all effects leading to virialization of the system.
We demonstrate explicitly in an appendix that velocity dispersion
terms arise naturally in the ``force'' equation (for the function $h
\equiv - v/{\dot a}x$), derived from the BBGKY hierarchy, and play the
same role as the function $S(\delta)$ in the fluid picture. This clearly
justifies the above procedure and shows that our approach could have a
somewhat larger domain of validity than might be expected from an analysis
based on the fluid picture.

\section{The virialization term}
We will now derive an approximate functional form for the virialization 
function from physically well-motivated arguments. 
If the virialization term is retained in equation (\ref{reqn}), we have
\begin{equation}
\label{theRequation}
{{d^2 R}\over {d t^2}}=-{{GM}\over {R^2}} - {{H^2 R} \over 3} S  
\end{equation} 
where $H=\dot a/a$. 
Let us first consider the late time behaviour of the system. When virialization 
occurs, it seems reasonable to
assume that  $R\rightarrow  {\rm constant} $  and $\dot{R} \rightarrow 0$. 
This implies that, for large density contrasts, 
\begin{equation}
S \approx  -{{3GM} \over {R^3 H^2}} \;\; \qquad(\delta \gg 1) 
\end{equation}

\noindent Using $H=\dot{a}/a=(2/3t)$, and equation  (\ref{deltadefn})
\begin{equation}
 S  \approx  -{{27GM t^2} \over {4 R^3}} = -{3 \over 2} (1 + \delta
)\approx -{3\over 2}\delta \;\; \qquad(\delta \gg 1)
\end{equation}
Thus, the ``virialization'' term tends to a value of ($ -3 \delta/2$) in the non-linear 
regime, when stable structures have formed. This asymptotic form for 
$S(\delta)$ is, however, insufficient to model its behaviour 
  over the  larger range of density contrast (especially the 
quasi-linear regime) which is of interest to us. Since $S(\delta)$ 
tends to the above asymptotic form at late times, the residual part, {\it i.e.} 
the part that remains after the asymptotic value has been subtracted away, 
can be expanded in a Taylor series in $(1 / \delta)$ without any loss of generality.
Retaining the first two terms of expansion, we write the complete virialization term as 
\begin{equation}
\label{netrotation2}
S(\delta)=-\frac{3}{2} (1+\delta) -\frac{A}{\delta}+\frac{B}{\delta^2}
+{\cal O}(\delta^{-3})
\end{equation}
\noindent Replacing for $S(\delta)$ in equation ({\ref{deltalog}), we obtain, 
for $\delta \gg 1$ \\
\begin{equation}
\label{new_dhdDeqn}
3h\delta \frac{dh_{\rm SC}}{d\delta} - h^2_{\rm SC} + \frac{h_{\rm SC}}{2} + 
\frac{1}{2}  = -\frac{A}{\delta}+\frac{B}{\delta^2}
\end{equation}
\noindent [It can be easily demonstrated that  the first order term in the 
Taylor series is alone insufficient  to model the turnaround behaviour of the 
$h$ function. We will hence include the next higher order term and use the 
form in equation (\ref{netrotation2}) for the virialization  term. The signs are 
chosen for future convenience, since it will turn out that both $A$ and $B$ are 
greater than zero.] In fact, for sufficiently large $\delta$, the 
evolution depends only on the combination $q\equiv(B/A^2)$. This is most
easily seen by rewriting equation (\ref{deltaequation}), replacing $S(\delta)$ 
with the above form. Taking the limit of 
large $\delta$, {\it i.e.} $\delta \gg 1$, and rescaling  $\delta$
to $\delta/A$, we  obtain \\
\begin{eqnarray}
\label{deltaequationscaled}
\frac{d^2\delta}{d b^2}+\frac{3}{2 b}\frac{d \delta}{d b}-\frac{4}{3 \; \delta} \left(
\frac{d \delta}{d b}\right)^2 \:&=&\:  -\frac{1}{a^2}+\frac{B}{A^2}\frac{1}{a^2\,\delta} 
\\
&=&-\frac{1}{a^2}+\frac{q}{a^2\delta}
\end{eqnarray}

\noindent From the form of the equation it is clear that the constants 
$A$ and $B$ occur in the combination $q=B/A^2$ and hence the non-linear
regime is modelled by a one parameter family for
 the virialization  term. 
\par
\noindent Equation (\ref{theRequation}) can be written as
\begin{equation}
\label{theRequation2}
\ddot{R}=-\frac{GM}{R^2}-\frac{4R}{27t^2} \left[ -\frac{27 GMt^2}{4 R^3}- \frac{A}{\delta}+ \frac{B}{\delta^2} \right]
\end{equation}
Using $\delta=9GMt^2/2R^3$ and $B=qA^2$ we may express equation (\ref{theRequation2}) 
completely in terms of $R$ and $t$. We now rescale $R$ and $t$ in the form 
$R=r_{vir}y(x)$ and $t=\beta x$, where $r_{vir}$ is the final virialised 
radius [{\it i.e.} $R \rightarrow r_{vir}$ for $t \rightarrow \infty$], where 
$\beta^2=(8/3^5) (A/GM) r_{vir}^3$, to obtain the following equation for $y(x)$\\
\begin{equation}
\label{thescaledeqn}
y''=\frac{y^4}{x^4} -\frac{27}{4} q \frac{y^7}{x^6}
\end{equation}
We can integrate this equation to find a form for $y_q(x)$ (where $y_q(x)$ is the function $y(x)$ for a specific value of $q$) using the physically motivated boundary conditions $y=1$  and $y'=0$ as $x \rightarrow \infty$, which is simply an expression of the fact that the system reaches the virial radius $r_{vir}$ and remains here thereafter.

\begin{figure}
\centering
\psfig{file=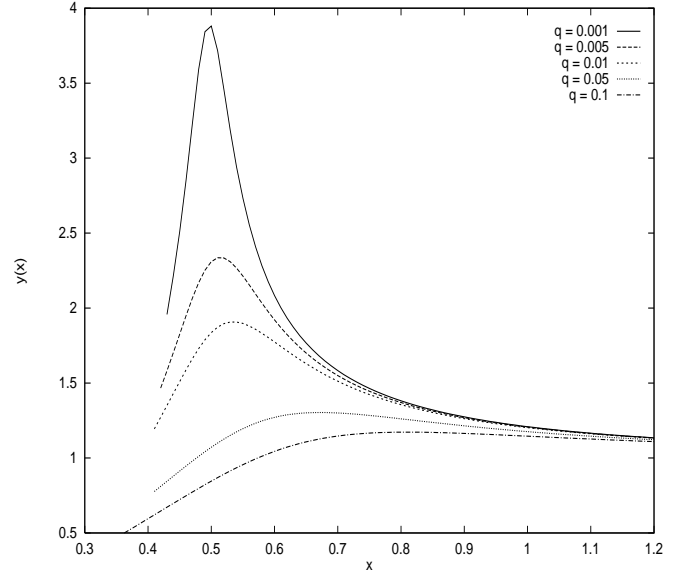,width=3.5truein,height=3.0truein,angle=-90}
\caption{The figure shows  the function $y_q(x)$ for some values of $q$. The x axis has scaled time, $x$ and the y axis is the scaled radius $y$.}
\label{figure1}
\end{figure}
\par
\noindent The results of numerical integration of this equation for a range of $q$ 
values are shown in figure (\ref{figure1}). 
As  expected on physical grounds,  the function has a maximum and gracefully decreases 
to unity for large values of $x$ [the behaviour of $y(x)$ near $x=0$ is irrelevant since the 
original equation is valid only for $\delta \geq 1$, at least]. For a given value of 
$q$, it is possible to find the value $x_c$ at which the function reaches its maximum, 
as well as the ratio $y_{max}=R_{max}/r_{vir}$. The time, $t_{max}$,  at which the 
system will reach the maximum radius is related to $x_c$ by the relation $t_{max}=
\beta x_c = t_0 (1+z_{max})^{-3/2}$, where $t_0=2/(3 H_0)$ is the present age of 
the universe and $z_{max}$ is the redshift at which the system turns around. 
Figure (\ref{figure2a}) shows the variation of $x_c$ and $y_{max}\equiv 
(R_{max}/r_{vir})$ for different values of $q$. The entire evolution of the 
system in the modified spherical collapse model (MSCM) can be expressed in terms of 

\begin{equation}
\label{MSCMsoln}
R(t)=r_{vir}\; y_q(t/\beta)
\end{equation} 
where $\beta=(t_0/x_c) (1+z_{max})^{-3/2}$.
\par
\noindent In SCM, the conventional value used for ($r_{vir}/R_{max}$) is ($1/2$), 
which is obtained by enforcing the virial condition that $\vert U \vert=2K$, where 
$U$ is the gravitational potential energy and $K$ is the kinetic energy. It must 
be kept in mind, however, that the ratio ($r_{vir}/R_{max}$) is not really 
constrained to be {\it precisely} ($1/2$) since the 
actual value will depend on the final density profile and the precise definitions
used for these radii. While we expect it to be around $0.5$, some
 amount of variation, say between 0.25 and 0.75, cannot be ruled out theoretically.
\begin{figure}
\centering
\psfig{file=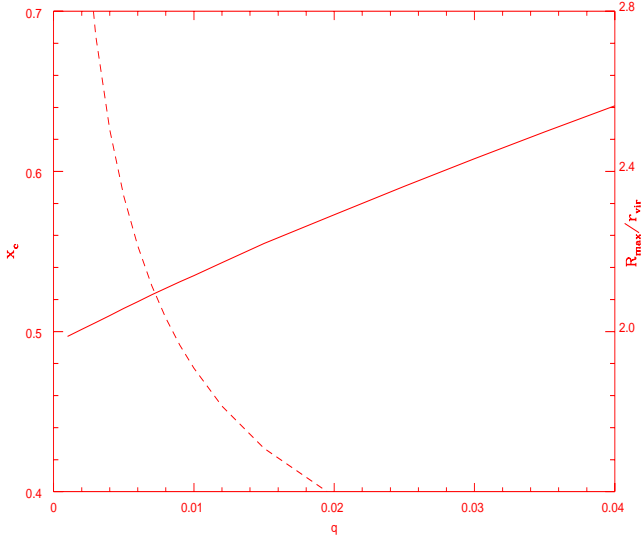,width=3.5truein,height=3.0truein,angle=-90}
\caption{The figure shows the parameters ($R_{max}/r_{vir}$) (broken line)  and $x_c$ 
(solid line) as a function of $q=B/A^2$. This clearly demonstrates that the single 
parameter description of the virialization term  is constrained by the value that is 
chosen for the ratio $r_{vir}/R_{max}$.}
\label{figure2a}
\end{figure}

\noindent Figure (\ref{figure2a}) shows the parameter ($R_{max}/r_{vir}$),   
plotted as a function of $q=B/A^2$ (dashed line),
obtained by numerical integration of  
equation (\ref{theRequation}) with the {\it ansatz\/}  (\ref{netrotation2}).
The solid line  gives the dependence of $x_c$ (or equivalently $t_{max}$) 
on the value of $q$. It can be seen that one can obtain a suitable value for 
the ($r_{vir}/R_{max}$) ratio by choosing a suitable value for $q$ and vice versa.   
Using equation (\ref{defh}) and the definition $\delta \propto t^2/R^3$, we obtain 
\begin{equation}
h_{\rm SC}(x)=1-\frac{3}{2} \frac{x}{y} \frac{d y}{d x}
\end{equation}
which gives the form of $h_{\rm SC} (x)$ for a given value of $q$; this, in turn, 
determines the function $y_q(x)$.
Since $\delta$ can be expressed in terms of $x$, $y$ and $x_c$ as $\delta=
(9 \pi^2/2 x_c^2) x^2/y^3$, this allows us to implicitly obtain a form for 
$h_{SC}(\delta)$, determined only by the value of $q$.
\begin{figure}
\centering
\psfig{file=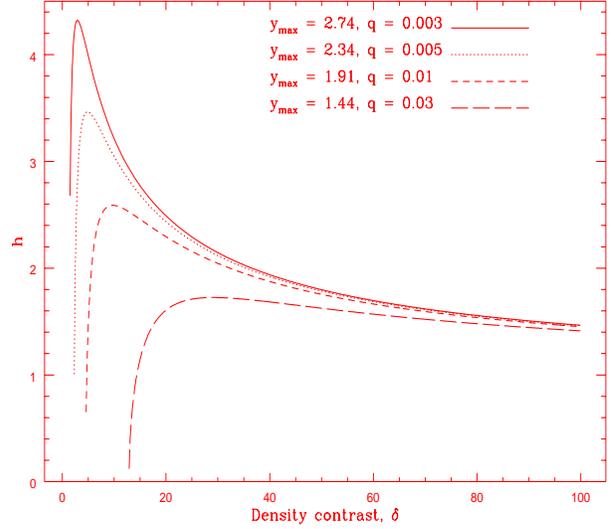,width=3.5truein,height=3.0truein,angle=-90}
\caption{The figure shows  the $h_{\rm SC}$ function, obtained for  various 
values of $q$. The values of $q$ and $y_{max} \equiv R_{max}/r_{vir}$ for 
the curves are indicated at the top right hand corner. (Further discussion in text) }
\label{figure2}
\end{figure}
\noindent Figure (\ref{figure2}) shows the  behaviour of $h_{SC}$
 functions obtained by integrating equation (\ref{dhdDeqn}) backwards,
 assuming that $h_{SC} \rightarrow 1$ as $\delta \rightarrow \infty$.
 It is seen that all the curves  have the same turnaround behaviour  
expected on the basis of the physical arguments presented in the earlier 
section. 
\par
\noindent If the functional form for  $h_{SC}$ -- determined, say, from N-body simulations -- is
used as a further constraint, we should be able to obtain the values of $q$. 
The major hurdle in attempting to do this is the fact that the available 
simulation results are given in terms of the averaged two point correlation 
function, $\bar\xi$, and the averaged pair velocity, $h(a,x)$, defined by 
\begin{equation}
\bar\xi=\frac{3}{r^3}\int_0^r \xi(x,a) x^2 dx \; ;\qquad h(a,x) =
-\frac{\left< v(a,x) \right>}{\dot{a} x}
\end{equation}
\noindent where the two-point correlation function $\xi$ is defined as 
the Fourier transform of the power spectrum, $P(k)$, of the distribution.
 The results published in the literature  assume that $h(a,x)$ 
depends on $a$ and $x$ only through $\bar\xi(a,x)$, that is, $h(a,x) 
\equiv h[\bar\xi(a,x)]$. This assumption has been invoked in several papers
in the past (See e.g. Hamilton et al. 1991, Nityananda $\&$ Padmanabhan
1994, Mo et al. 1995, Padmanabhan 1996a, Padmanabhan $\&$ Engineer 1998)
and seems to be validated by numerical simulations. The
fitting formula for $h(\bar\xi)$ can be obtained from related fitting formulas 
available in the literature (e.g. Hamilton et al. 1991). These are, however, statistical 
quantities and are not well 
defined  for an isolated overdense region. Hence we have to first make the 
correspondence between $h_{SC}(\delta)$ and $h(\bar\xi)$, which we do as follows. 
\par
\noindent It is possible to show by standard arguments 
(Nityananda $\&$ Padmanabhan 1994) that \\
\begin{equation}
\label{paddyh}
{{d \bar\xi}\over {d\alpha}} = 3h(1 + \bar\xi)
\end{equation}
\noindent that is,\\
\begin{equation}
\label{hxieqn}
{{d D}\over {d\alpha}} = 3h
\end{equation}
\noindent where $D = {\rm ln} (1+\bar\xi)$ and $\alpha={\rm ln}\;a$. Equation (\ref{hxieqn}) 
is very similar to equation (\ref{defh}), which defines the function 
$h_{\rm SC}(\delta)$, except for the different definitions of $D$ and 
$D_{\rm SC}$ 
in terms of $\bar\xi$ and $\delta$ respectively. This suggests  that one 
can obtain a relation between $h_{\rm SC}(\delta)$ and $h(\bar\xi)$ 
by  relating  the density contrast $\delta$ of an isolated 
spherical region to the two-point correlation function $\bar\xi$ 
averaged over the distribution at the same scale. We essentially need to find 
a mapping between $\bar\xi$ and $\delta$ which is valid in a statistical sense.
\par
\noindent Gravitational clustering is known to have three regimes in its
growing phase, usually called ``linear'', ``quasi-linear'' and ``non-linear'' 
respectively. The three regimes may be characterized by values of 
density contrast as  
$\delta \ll 1$ in the linear regime, $1<\delta < 100$ in the quasi-linear regime 
and  $100<\delta$ in the non-linear regime. The three regimes have different rates 
of growth for various quantities of interest such as $\delta$, $\xi$ and so on. 
In the linear regime, it is well known that
the density contrast grows proportional to the scale factor, $a$.
 This implies that the power spectrum, $P(k) \equiv |\delta_k|^2$ (where
$\delta_k$ is the Fourier mode corresponding to $\delta(x)$), grows as
$a^2$. Consequently, $\bar\xi$, which is related to $P(k)$ via a Fourier
transform, also grows as $a^2$, {\it i.e.} as the square of the density
contrast. In the quasi-linear and non-linear regimes, the density contrast 
does not grow linearly with the scale factor and the relation between 
$\delta$ and $\bar\xi$ is not so clearly defined. The quasi-linear regime 
may be loosely construed as the interval of time during which the high peaks
of the initial Gaussian random field have collapsed, although mergers
of structures have not yet begun to play an important role.
(This idea was used in  Padmanabhan (1996a) to model the non-linear scaling relations
successfully). If we consider
a length scale smaller than the size of the collapsed objects, the dominant 
contribution to $\bar\xi$ (at this scale)
arises from the density profiles centered on the collapsed
peaks. Using the relation \\
\begin{equation}
\rho \simeq  \rho_b\left( 1 + \bar\xi \right)
\end{equation}
\noindent for density profiles around high peaks, one can see that $\bar\xi \propto \delta$ in this regime.
 In the non-linear regime, $\delta$ and $\bar\xi$ 
have the forms $\delta(a,x)=a^3F(ax)$, $\bar\xi=a^3 G(ax)$, where $x$ is a 
co-moving and $r=ax$ is a proper coordinate. When the system is described by Lagrangian 
coordinates (which correspond to 
proper coordinates $r=ax$, {\it i.e.\/}  at constant $r$), $\bar\xi$ is proportional 
to $\delta$. Thus, the relation $\bar\xi \propto \delta$
appears to be satisfied in all regimes, except at the very linear 
end. Since we are only interested in the $\delta >1$ range, we use
$\bar\xi \approx \delta$
\noindent and compare equations (\ref{paddyh}) and (\ref{defh}) to identify \\
\begin{equation}
h_{\rm SC}(\delta) \approx h(\bar\xi) 
\end{equation}
It is now straightforward to choose the value of $q$ such that the
known fitting function for the $h$ function is reproduced as closely as possible.
We use the original function given by Hamilton et al. (1991) to obtain the following 
expression for $h(\bar\xi)$:
\begin{equation}
\label{hamiltonfit}
h(\bar\xi)=\frac{2}{3} \left(\frac{d \ln {\cal V}(\bar\xi)}{d \ln (1+\bar\xi)}\right)^{-1}
\end{equation} 
where ${\cal V}(\bar\xi)$ is given by the fitting function 
\begin{equation}
\label{Vhamilton}
{\cal V}(\bar\xi)=\bar\xi \left( \frac{1+0.0158\;{\bar\xi}^2+0.000115\;{\bar\xi}^3}{1+0.926\;{\bar\xi}^2-0.0743\;{\bar\xi}^3+0.0156\;{\bar\xi}^4}\right)^{1/3} \end{equation} 

\begin{figure}
\centering
\psfig{file=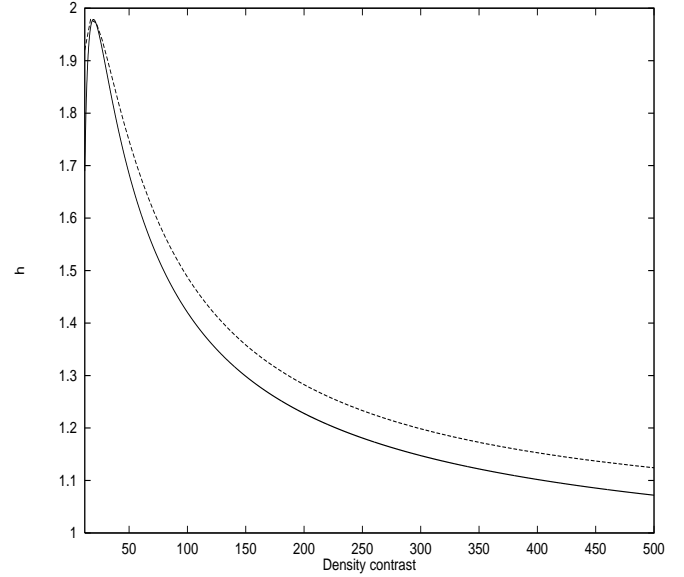,width=3.5truein,height=3.0truein,angle=-90}
\caption{The figure shows the best fit curve for  the $h$ function (dashed line) to the simulation data  (solid line). The simulation results are obtained from Hamilton et al. (1991) and the fit is obtained  by adjusting the value of $q$ parameter until the curves coincide. }
\label{figure3}
\end{figure}

\noindent Figure (\ref{figure3}) shows the simulation data 
represented by the fit (solid line) \cite{ham} and the best fit (dashed line), 
obtained in our model, for $q\simeq 0.02$. We note that the fit is better 
than 5\% for all values of density contrast $\delta  \ge 15$. The change in the 
fit is very marginal if one imposes the boundary condition $h(\delta) 
\rightarrow 1$ for $\delta \gg 1$, instead of constraining the curves to match 
at their peaks (for example, the change in the peak height is $\sim 1 \%$, 
if we impose the above condition at $\delta = 10000$). \\

\begin{figure}
\centering
\psfig{file=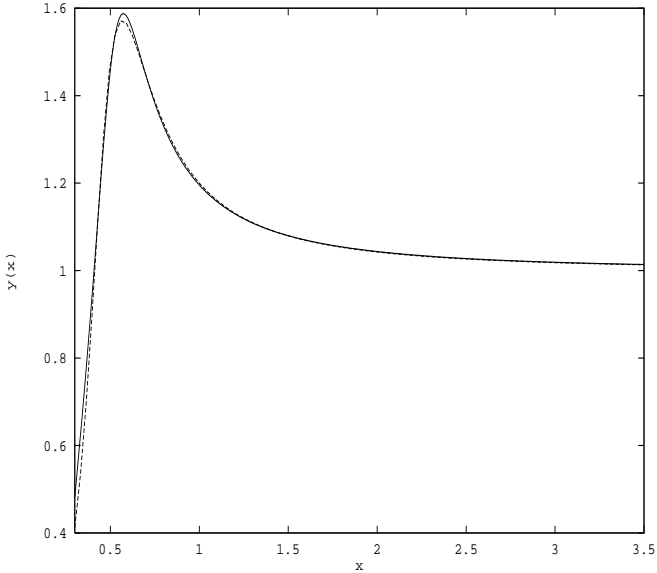,width=3.5truein,height=3.0truein,angle=-90}
\caption{The figure shows a plot of the scaled radius of the shell $y_q$ as a function of scaled time $x$ (solid line) and the fitting formula $y_q=(x+ax^3+bx^5)/(1+cx^3+bx^5)$, with $a=-3.6$, $b=53$ and $c=-12$ (dashed line) (See text for discussion) }
\label{figure5}
\end{figure}
\noindent Figure (\ref{figure5}) shows the plot of scaled radius $y_q(x)$ vs $x$, 
obtained by integrating equation(\ref{thescaledeqn}), with $q=0.02$.
The figure also shows an accurate fit (dashed line) to this solution of the form
\begin{equation}
\label{yqfit}
y_q(x)=\frac{x+a x^3+b x^5}{1+c x^3+b x^5}
\end{equation} 
with $a=-3.6$, $b=53$ and $c=-12$. This fit, along with values for $r_{vir}$ 
and  $z_{max}$, completely specifies our model through equation (\ref{MSCMsoln}).
It can be observed  that ($r_{vir}/R_{max}$) is approximately $0.65$.
It is  interesting to note that the value  obtained for the 
($r_{vir}/R_{max}$) ratio is not very widely off the usual value of $0.5$ used 
in the standard spherical collapse model, {\it  in spite of the fact that no 
constraint was imposed on this value, {\it ab initio},  in arriving at 
this result.\/} Part of this deviation {\it may} also originate in the fit 
which has been used for $h(\bar\xi)$; Hamilton et al. (1991) noticed that 
objects virialised at $ R_{max}/r_{vir} \sim 1.8$, instead of 2, in their 
simulations. \\
\par 
\begin{figure}
\centering
\psfig{file=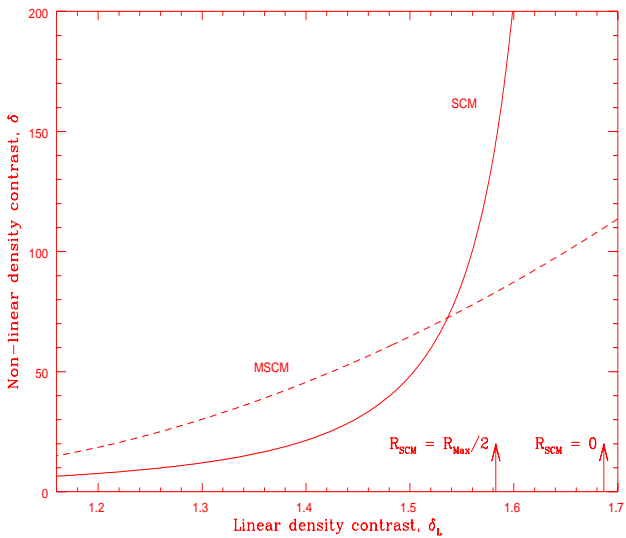,width=3.5truein,height=3.0truein,angle=-90}
\caption{The figure shows the non-linear density contrast in the SCM 
(solid line) and in the modified SCM (dashed line), plotted against 
the linearly extrapolated density contrast $\delta_L$ (discussion in text). }
\label{figure6}
\end{figure}

\noindent  Finally, figure (\ref{figure6}) compares the non-linear  density
contrast  in the modified SCM (dashed line) with that in the standard SCM 
(solid line), by plotting both against the linearly extrapolated density contrast,
$\delta_L$. 
It can be seen (for a given system with the same $z_{max}$ and 
$r_{vir}$) that, at the epoch where the standard SCM model has a
singular behaviour ($\delta_L \sim 1.686$), our model has a smooth 
behaviour with $\delta \approx 110$ (the value is not very sensitive 
to the exact value of $q$). This is not widely off from the value 
usually obtained from the {\it ad hoc}  procedure applied in the standard 
spherical collapse model. In a way, this explains the unreasonable
effectiveness of standard SCM in the study of non-linear clustering. 
\par
\noindent As mentioned earlier, deviations from spherical symmetry
are expected to be small at early epochs and to grow as the
system evolves. One would thus expect the two curves of figure (\ref{figure6})
to approach each other as $ \delta \rightarrow 1$ (from above).
Further, the curves should overlay in the linear regime ($\delta_L
\ll 1$). It can be seen from the figure that the 2 curves {\it do}
approach each other as $\delta_L$ reduces towards unity. However, 
the MSCM has been obtained using a Taylor expansion in $(1/\delta)$; 
it is clearly {\it not} applicable for $\delta \ll 1$. Further,
the region $\delta > 15$ has been used to fit the function $h(\delta)$ 
to the data of Hamilton et al. (1991). Hence, one cannot compare the 
curves in the linear regime.
\par
\noindent Figure (\ref{figure6}) also shows a comparison between the 
standard SCM and the MSCM in terms of $\delta$ values in the MSCM at 
two important epochs, indicated by vertical arrows. (i) 
When $R = R_{max}/2$ in the SCM, {\it i.e.} the epoch at which 
the SCM {\it virializes}, $\delta({\rm MSCM}) \sim 83$.
(ii) When the SCM hits the singularity, ($\delta_L \sim 1.6865$), 
$\delta({\rm MSCM}) \sim 110$.\\
\par
\noindent We note, finally, that figure (\ref{figure6}), which shows the effects  
of evolution as a mapping from linear to non-linear density contrasts, 
contains a subtle implicit assumption regarding the definition of the 
non-linear density contrast. The radius $R$ of a system is not a 
rigorously defined quantity in the absence of spherical symmetry, 
and obviously, any argument involving `virialization' precludes 
strict spherical 
symmetry. It is, however, a conventional 
practice to define the `radius', $R$, 
even for a virialized system without strict spherical symmetry. 
For example, this approach is used to define the density contrast at 
`virialization' (which has the value, $\delta_{vir} \approx \; 200$) in 
the standard SCM. In our model we have an explicit equation for $R$; once $R$ 
and $M$ are given, the non-linear density contrast is a well-defined quantity.\\
\par

\section{Results and Conclusions}

\noindent In this paper, we have shown how the Taylor expansion of a 
term in the equation for the evolution of the  
density contrast, $\delta$, in inverse powers of $\delta$, allows us to have a 
more realistic picture of spherical collapse, which is free from arbitrary 
abrupt ``virialization'' arguments. Beginning from a well motivated {\it ansatz\/}  
for the dependence of the ``virialization'' term on the density contrast we have shown 
that a spherical collapse model will gracefully turn around and collapse to a 
constant radius with  $\delta \sim 110$ at the same epoch when the standard 
model  reaches  a singularity. Figure (\ref{figure5}) shows clearly that the singularity  
 is avoided in our model due to the enhancement of deviations from spherical symmetry, 
and consequent generation of strong non-radial motions.
\par
\noindent We derive an approximate functional form for the virialization  term 
starting from 
the physically reasonable assumption that the system reaches a constant radius. This 
assumption allows us to derive an asymptotic form for the virialization term, with the 
residual part  adequately expressed by keeping only the first and second order terms 
in a Taylor series in ($1/\delta$).  It is shown that there exists a scaling relation 
between the coefficients of the first and second order terms, essentially reducing 
the virialization term to a one parameter family of models. 
\par
\noindent The form of the $h$ function published in the literature, along with 
a tentative  mapping from $\delta$ to $\bar\xi$, in the 
non-linear and quasi-linear regimes, allow us to further constrain our model, 
bringing it in concordance with the available simulation results. Further, it is 
shown that this form for the virialization  term  is sufficient to model 
the turnaround behaviour of the spherical shell and leads to a reasonable 
numerical value for density contrast at collapse. 
\par
\noindent There are several new avenues suggested by this work which we plan to 
pursue in the future. \\
\noindent (i) The assumption $h_{\rm SC} \rightarrow 1$, $R \rightarrow r_{vir}$ is 
equivalent to ``stable clustering'', in terms of the statistical behaviour. 
Since stable clustering has so far not been proved conclusively in simulations and 
is often questioned, it would be interesting to see the effect of changing this 
constraint to $h \rightarrow {\rm constant}$ for $t \rightarrow \infty.$

\noindent (ii) The technique of Taylor series expansion in ($1/\delta$) seems 
to hold promise. It would be interesting to try such an attempt with the 
original fluid equations and (possibly) with more general descriptions.

\noindent (iii) It must be stressed that we used the $\delta-\bar\xi$ mapping
--- possibly the weakest part of our analysis, conceptually --- only to fix 
a value of $q$. We could have used some high resolution simulations to actually 
study the evolution of a realistic overdense region. We conjecture that such 
an analysis will give results in conformity with those obtained here. We plan 
to investigate this --- thus eliminating our reliance on the $\delta -\bar\xi$ 
mapping --- in a future work.

\noindent (iv) Finally, the curves of figure (6) can be used to describe the spatial
distribution of virialised haloes (see, for example, Mo \&
White 1996; Sheth 1998). It would be interesting to investigate
how things change when the MSCM is used in place of the standard
spherical collapse model.\\

\section{Appendix}
\noindent The zeroth and first moments of the 2$^{\rm nd}$ BBGKY equation 
(Ruamsuwan \& Fry (1992), equations (22) and (23)) can be combined to obtain the 
following equation for the dimensionless function $h = -v/{\dot a}x$ \cite{kanekar} \\
\begin{eqnarray}
\label{hevolution}
3h(1 + \overline\xi) {{dh}\over {d\overline\xi}} + {h \over 2} - h^2 -
{{3\overline\xi} \over F} - {{9MQe^{-X}}\over {4\pi F}} \quad =  \qquad \quad \nonumber \\
 \qquad\qquad \qquad \qquad \qquad h^2_{\parallel} {\Big (} 4 + 
{{{\partial \mbox{ln}\;\;  F}\over{\partial{X}}}} 
 {\Big )} + {{{\partial h^2_{\parallel}}\over{\partial{X}}}} -  2 h^2_{\perp} 
\end{eqnarray}
\noindent where we have used the {\it ansatz}, $h \equiv h({\overline\xi})$ (Hamilton et al. 1991; 
Nityananda \& Padmanabhan 1995). In the above, we have defined \\
\begin{equation}
\label{barxi}
\overline{\xi}(x,a) = {3 \over {x^3}}\int_0^x dx \xi(x,a) x^2 $$
\end{equation}
\begin{equation}
F = {{{\partial {\overline\xi}}\over{\partial{X}}}} + 3(1 + \overline\xi)\:\; 
, \:\; X = {\rm ln}\; x
\end{equation}
\begin{equation}
h^2_{\parallel} = {{\Pi} \over {{\dot a}^2x^2}} \:\;,\:\; 
h^2_{\perp} = {{\Sigma} \over {{\dot a}^2x^2}}
\end{equation}
\noindent where $\Pi$ and $\Sigma$ are parallel and perpendicular peculiar 
velocity dispersions (Ruamsuwan \& Fry 1992; Kanekar 1999). Finally, we have assumed 
that the 3-point correlation function has the hierarchical form (Davis \& Peebles 1977;
 Ruamsuwan \& Fry 1992) \\
\begin{equation}
\zeta_{123} = Q(\xi_{12}\xi_{13} + \xi_{13}\xi_{23} + \xi_{12}\xi_{23})
\end{equation}
\noindent and defined \\
\begin{equation}
M = \int d^3z{\Big [} \xi(x) + \xi(z){\Big ]}\xi({\bf z}-{\bf x})
{{\mbox{cos}\;\theta} \over {z^2}}
\end{equation}
\noindent In the non-linear regime, $\overline\xi \gg 1$, the stable clustering ansatz 
yields a scale-invariant power-law behaviour for $\overline\xi$ \cite{DP}, 
with $\overline\xi \propto a^{(3-\gamma)}x^{-\gamma}$, if $h \rightarrow 1$ as 
$\overline\xi \rightarrow \infty$. In this limit, we have \\
\begin{equation}
F = (3-\gamma)\overline\xi + 3
\end{equation}
\noindent and \\
\begin{equation}
{{{\partial \mbox{ln} \; F}\over{\partial{X}}}} = -\gamma
{\Big[} 1 + ({3 \over {3 - \gamma}})({1 \over \overline\xi}) {\Big]}^{-1}$$
\end{equation}
\noindent Further, we can write \cite{YG} $M = M'x\overline\xi^2$, where $M'$ 
is a constant. Thus, equation (\ref{hevolution}) reduces, in the non-linear regime,
to \\
\begin{eqnarray}
3h\overline\xi \frac{dh}{d\overline\xi} - h^2 + \frac{h}{2} - 
\frac{3}{3-\gamma} \left[ \frac{3M'}{4\pi} \left\{ \overline\xi - \frac{3}{3-\gamma}\right\}
+ 1 \right]   = \nonumber \\
\left[ \left(4 - \gamma\right) + \frac{3\gamma}{(3-\gamma)\overline\xi}\right]h^2_\parallel
+ {{{\partial h^2_{\parallel}}\over{\partial{X}}}} -  2 h^2_{\perp} + 
{\cal O}\left(\frac{1}{\overline\xi}\right)
\end{eqnarray}
\noindent where we have retained terms upto order constant in $\overline\xi$. We now 
assume that $h^2_\parallel$ and $h^2_\perp$ are functions of $\overline\xi$ alone, to first 
order. This yields \\
\begin{eqnarray}
\label{geqn}
3h\overline\xi \frac{dh}{d\overline\xi} -h^2 + \frac{h}{2} - 
\frac{3}{3-\gamma} \left[ \frac{3M'}{4\pi} \left\{ \overline\xi - \frac{3}{3-\gamma}\right\}
+ 1 \right] = \nonumber \\
G(\overline\xi) + {\cal O}\left(\frac{1}{\overline\xi}\right) 
\end{eqnarray}
\noindent with \\
\begin{equation}
G(\overline\xi) = \left[ \left(4 - \gamma\right) + \frac{3\gamma}{(3-\gamma)
\overline\xi}\right] h^2_\parallel(\overline\xi)
- \gamma\overline\xi \frac{d h^2_\parallel}{d \overline\xi}-  2 h^2_{\perp}(\overline\xi)
\end{equation}
\noindent Clearly, if $h \rightarrow 1$ as $\overline\xi \rightarrow \infty$, we must have \\
\begin{eqnarray}
G(\overline\xi) = - \frac{3}{3-\gamma} \left[ \frac{3M'}{4\pi} \left\{ \overline\xi - 
\frac{3}{3-\gamma}\right\} + 1 \right]  - \frac{1}{2} + {\cal O}\left(\frac{1}{\overline\xi}
\right)
\end{eqnarray}
\noindent {\it i.e.} $G(\overline\xi) \approx - {9M'\overline\xi}/{4\pi(3-\gamma)}$ 
for $\overline\xi \gg 1 $. \\
\noindent Since $G(\overline\xi)$ tends to the above asymptote at late times, the residual 
part can be expanded in a Taylor series in $ 1/\overline\xi$. Retaining the first two terms 
of the expansion in equation (\ref{geqn}), we obtain \\
\begin{equation}
3h\overline\xi \frac{dh}{d\overline\xi} -h^2 + \frac{h}{2} + \frac{1}{2} = 
\frac{A}{\overline\xi}+\frac{B}{\overline\xi^2} +{\cal O}(\overline\xi^{-3})
\end{equation}
\noindent This is exactly the same as equation (\ref{new_dhdDeqn}), with $\overline\xi$ 
replacing $\delta$. $G(\overline\xi)$ thus plays the same role as $S(\delta)$ in the 
stabilising of the system against collapse. This clearly implies that the 
velocity dispersion terms, $h^2_\parallel$ and $h^2_\perp$, will contribute to the 
support term; we are hence 
justified in writing the virialization  term in the more general form \\
\begin{equation}
S = a^2\left(\sigma^2 - 2\Omega^2\right) + f(a,x)
\end{equation}
where $f(a,x)$ contains contributions from effects arising from shell crossing, multi-streaming, 
etc.

\section{Acknowledgments}
\noindent  It is a pleasure to thank Ravi Sheth for detailed comments on an 
earlier version of this paper. SE would like to acknowledge CSIR for 
support during the course of this work. The research work of one of the 
authors (TP) was partly supported by the INDO-FRENCH Centre
for Promotion of Advanced Research under grant contract No 1710-1.\\

\end{document}